\documentstyle[aps,pre]{revtex}
\begin{document}
\draft
\title{Transversal inhomogeneities in dilute vibrofluidized granular fluids}
\author{J. Javier Brey, M.J. Ruiz--Montero, F. Moreno, and R. Garc\'{\i}a-Rojo}
\address{F\'{\i}sica Te\'{o}rica, Universidad de Sevilla, Apartado de
 Correos 1065, E-41080 Sevilla, Spain}
\date{today}

\maketitle

\begin{abstract}
The spontaneous symmetry breaking taking place in the direction perpendicular
to the energy flux in a dilute vibrofluidized granular system is investigated,
using both a hydrodynamic description and simulation
methods. The latter include molecular dynamics and direct Monte Carlo
simulation of the Boltzmann equation. A marginal stability analysis of the
hydrodynamic equations, carried out in the WKB approximation, is shown to be
in good agreement with the simulation results. The shape of the hydrodynamic
profiles beyond the bifurcation is discussed.

\end{abstract}

\section{Introduction}
\label{s1}
Granular materials are assemblies of macroscopic particles dissipating their
energy through inelastic collisions \cite{JNyB96}. They exhibit a very rich
phenomenology that is only partially understood. One of the most 
peculiar behaviors of granular systems, which has attracted 
a lot of attention in the last years,
is their tendency to spontaneously develop strong spatial inhomogeneities. In 
many different situations, the density shows a sharp profile that is not
induced by the boundary conditions. This phenomenon is often referred to
as  a clustering effect \cite{GyZ93}, since high density regions coexist in
the system with regions where the density is very low.  

In vibrated granular systems, clustering effects show up in many cases as a
spontaneous symmetry breaking  in the direction parallel to the vibrating
wall. Consider a gas enclosed in a box that is being supplied energy through
a vibrating  wall located at $x=0$. There are no other external forces
acting on the system. The box is divided into two
equal compartments by a wall normal to the $x$ axis starting at a 
certain height. At sufficiently low average density, the hydrodynamic 
fields are symmetric to both sides of the partition, but above a 
critical average density, which depends on the
value of the restitution coefficient, an asymmetry in the number of particles
at each side of the container occurs \cite{BMGyR02}. This asymmetry has been
shown to be associated with a bifurcation of the solution of the hydrodynamic
equations describing the state of the system. 

A similar symmetry breaking has been observed in a system in presence of
a gravitational force acting in the $x$ direction. In this case, the system
is unbounded for $x>0$, and the partition has a hole at a certain height.
Again, an asymmetry in the number of particles in the two compartments 
develops if
a control parameter, dependent on the amplitude of the vibration and the
degree of inelasticity, is larger than a critical value \cite{SyN96,Eg99}.

Symmetry breaking in the direction parallel to the vibrating wall has
also been observed in systems without any partition of the container.
Sunthar and Kumaran \cite{SyK01} have reported molecular dynamics simulation
results showing the presence of convection rolls and phase separation into
coexisting dense and dilute regions in a granular system in presence of
gravity. The phase separation takes place on the surface of the vibrating
wall and, as already indicated,
in the direction perpendicular to the energy flux. No theoretical explanation
for this phenomenon is provided in ref.\ \cite{SyK01}, although the
effect of the different parameters controlling the behavior of the
system is discussed in detail, on the basis of the simulation results.        
For a two-dimensional closed system in absence of gravity, a 
transversal continuous spontaneous symmetry
breaking has also been predicted  \cite{LMyS02}. Not at all surprisingly, the
gradients are now sharper next to the elastic wall, opposite to the energy
source. This is consistent with the positions that must have the holes 
in systems with a separating wall in order to observe the symmetry breaking, 
with and  without gravitational field acting on the system. The work 
by Livne {\em et al.} \cite{LMyS02} is restricted to the
nearly elastic limit and it is based on a numerical marginal stability
analysis of the hydrodynamic equations. The predictions of this analysis are
compared with numerical solutions of the own hydrodynamic equations with
the appropriate boundary conditions.

In this paper, the bifurcation predicted in ref.\  \cite{LMyS02} will be
considered again. There are several reasons for that. Firstly, hydrodynamic
equations derived from the Boltzmann equation for smooth inelastic hard disks
and valid, in principle, for
arbitrary inelasticity will be used, then somehow extending the previous
results. Secondly, instead of a numerical analysis of the stability of
the solutions of the hydrodynamic equations, an analytical study, based on
the WKB approximation, will be presented here. One of the main advantages of 
this approach is that the dimensionless control parameter governing
the bifurcation phenomenon is clearly identified. A third motivation for the
present work is to report simulation results, both from molecular
dynamics and also from Monte Carlo simulation of the Boltzmann equation,
showing the existence of the predicted transition. Since these simulation
techniques do not contain any externally introduced hydrodynamic concept,
they provide a direct proof of the existence of the continuous symmetry
breaking, and a test of the theoretical predictions. Attention will 
be also paid to the form of the hydrodynamic
profiles beyond the bifurcation. This leads to a deeper understanding of the
development of the instability. In any case, it is clear that the work in ref.\
\cite{LMyS02} opened the way to more systematic investigations of this
instability, like the one in this paper.

The plan of the paper is as follows. In Sec.\ \ref{s2}, the hydrodynamic
description of the one-dimensional state of a low density vibrofluidized
granular gas will be briefly summarized. The analytical expressions of the
hydrodynamic profiles are given. This state is the starting point for the
marginal stability analysis developed in Sec.\  \ref{s3}. Linearization
of the hydrodynamic equations around the one-dimensional state leads to
a second order linear differential equation. The WKB solution of the
closed problem posed by this equation and the corresponding boundary
conditions is built up. This requires to consider three different cases,
depending on the value of the parameters characterizing the system.
From the WKB solution, the marginal stability curve follows easily.
Simulation results are presented and compared with the theoretical
predictions in Sec.\ \ref{s4}, where an order parameter characterizing 
the transition is defined. A good agreement is found. 
The last section contains some final remarks, as well as the comparison 
of the results derived here with those in ref.\
\cite{LMyS02} in the common range of applicability, namely nearly elastic
collisions and very low density.

\section{Basic equations and the reference state}
\label{s2}
For a steady state without macroscopic flows, the balance equations for a
two-dimensional gas of smooth inelastic hard disks of mass $m$ and diameter 
$\sigma$ have the form
\begin{equation}
\label{2.1}
\nabla \cdot {\sf P} =0,
\end{equation}
\begin{equation}
\label{2.2}
 \nabla \cdot {\bf q} +n T \zeta=0,
\end{equation}
where $n$ and $T$ are the number density and granular temperature (with
Boltzmann's constant set equal to unity), 
respectively. In the low density limit, for a gas described by the inelastic
Boltzmann equation, and to lowest order in the gradients (Navier-Stokes order),
the pressure tensor ${\sf P}$ and heat flux ${\bf q}$, for the steady
state under consideration are given by
\cite{BDKyS98,ByC01}
\begin {equation}
\label{2.3}
{\sf P}=p {\sf I},
\end{equation}
\begin{equation}
\label{2.4}
{\bf q}=-\kappa \nabla T- \mu \nabla n,
\end{equation}
${\sf I}$ being the unit tensor, $p=nT$ the hydrodynamic pressure, $\kappa$ 
the heat conductivity, and $\mu$ a transport coefficient that has no analogue 
in the elastic case. These transport coefficients are proportional to the 
elastic heat conductivity $\kappa_{0}(T)$,
\begin{equation}
\label{2.5}
\kappa= \kappa^{*}(\alpha) \kappa_{0}(T), \quad
\mu=\mu^{*}(\alpha) \frac{T \kappa_{0}(T)}{n},
\end{equation}
\begin{equation}
\label{2.6}
\kappa_{0}(T)=\frac{2}{ \sigma} \left( \frac{T}{\pi m} \right)^{1/2}.
\end{equation}
Finally, the cooling rate in the same approximation is related with the 
elastic shear viscosity $\eta_{0}$ by
\begin{equation}
\label{2.7}
\zeta=\zeta^{*}(\alpha) \frac{p}{\eta_{0}},
\end{equation}
\begin{equation}
\label{2.8}
\eta_{0}(T)=\frac{1}{2 \sigma} \left( \frac{mT}{\pi} \right)^{1/2}.
\end{equation}
The functions $\kappa^{*}$, $\mu^{*}$, and $\zeta^{*}$ only 
depend on the constant
coefficient of normal restitution $\alpha$ characterizing the inelasticity 
of collisions. Their explicit expressions are given in refs. \cite{ByC01} and 
\cite{BRyM00}, and will be not reproduced here. 
By using the above expressions we get from Eqs. (\ref{2.1}) and (\ref{2.2}),
\begin{equation}
\label{2.9}
\frac{\partial p}{\partial x} =\frac{\partial p}{\partial y}=0,
\end{equation}
\begin{equation}
\label{2.10}
\left( \kappa^{*}-\mu^{\ast} \right) \left[ 
\frac{\partial}{\partial x} \left( \kappa_{0} \frac{\partial T}{\partial x} 
\right)
+\frac{\partial}{\partial y} \left( \kappa_{0} \frac{\partial T}{\partial y} 
\right) \right] -\zeta^{*} \frac{p^{2}}{\eta_{0}}=0.
\end{equation}
Next, we specify the boundary conditions. We consider that the system has $N$ 
particles enclosed
in a rectangular box with dimensions $L_{x}$ and $L_{y}$. The wall located at 
$x=0$ is vibrating and, therefore, supplies energy to the system. This energy
is needed in order to keep and sustain a fluidized steady state. The
other three walls, located at $x=L_{x}$, $y=0$, and $y=L_{y}$, respectively,
are at rest. For the sake of simplicity, collisions of particles with all 
four walls are assumed to be elastic. Then, the mathematical boundary 
conditions to be imposed are:
\begin{equation}
\label{2.11}
\left( \frac{\partial T}{\partial x} \right)_{x=L_{x}}=
\left( \frac{\partial T}{\partial y} \right)_{y=0}=
\left( \frac{\partial T}{\partial y} \right)_{y=L_{y}}=0,
\end{equation}
\begin{equation}
\label{2.12}
-\left[ \kappa^{*}(\alpha)-\mu^{*}(\alpha) \right] \left[ \kappa_{0}(T)
\frac{\partial T}{\partial x} \right]_{x=0}=Q.
\end{equation}
Equations (\ref{2.11}) express that the heat flux must vanish at the inmobile
walls, while Eq.\ (\ref{2.12}) is the energy balance at the vibrating wall.
The quantity $Q$ is the rate of energy input through this wall per unit
of length. Its calculation in terms of the parameters defining the motion of
the wall has been addressed in several works. Here we will consider the 
simplest
possibility, namely that the wall moves in a sawtooth manner with velocity 
$v_{b}$ \cite{BRyM00,McyB97,McyL98}. This is a good approximation to more 
realistic motions, as long as the characteristic frequency of vibration of 
the wall be much larger than the collision rate of the gas molecules in its 
vicinity. In addition, it will be assumed that the amplitude of the 
vibration is much smaller than the mean free path of the particles 
of the gas next to it, so collective motions in the system 
are not being generated. Under the above conditions, it is \cite{McyB97}
\begin{equation}
\label{2.13}
Q=p v_{b}.
\end{equation}
The closed mathematical problem defined by Eqs. (\ref{2.9})--(\ref{2.13}) 
admits a $y$-independent solution $T=T_{R}(x)$ that has been 
discussed in detail
in ref.\ \cite{BRyM00}. The existence of this one-dimensional solution had 
been previously noticed
by Grossman {\em et al.} \cite{GZyB97}. In the following, we will 
refer to this solution as the reference state and its properties will be 
characterized with a subindex $R$. The reference temperature profile is
\begin{equation}
\label{2.14}
T_{R}(\xi_{x})=T_{0,R} \left[ \frac{\cosh (\xi_{x}^{*}-\xi_{x})}
{\cosh \xi_{x}^{*}}
\right]^{2},
\end{equation}
where $\xi_{x}$ is a dimensionless scaled length defined by
\begin{equation}
\label{2.15}
\xi_{x}=\sqrt{a(\alpha)} \int_{0}^{x} \frac{dx}{\lambda_{R}(x)}\, ,
\end{equation}
with $\lambda_{R}(x)$ being the local mean free path,
\begin{equation}
\label{2.16}
\lambda_{R}(x)=\frac{1}{2 \sqrt{2} \sigma n_{R}(x)} \, ,
\end{equation}
and
\begin{equation}
\label{2.17}
a(\alpha)=\frac{\pi \zeta^{*}}{16(\kappa^{*}-\mu^{*})}\, . 
\end{equation}
Moreover, $\xi_{x}^{*}$ is the value of $\xi_{x}$ for $x=L_{x}$, i.e.,
\begin{equation}
\label{2.18}
\xi_{x}^{*}=2\sqrt{2 a(\alpha)}\,  \frac{\sigma N}{L_{y}}.
\end{equation}
Therefore, $\xi_{x}^{*}$ is proportional to the number of monolayers of 
particles perpendicular to the $x$-axis at rest, $\sigma N /L_{y}$. Finally,
the uniform pressure of the reference state is
\begin{equation}
\label{2.19}
p_{R}= \frac{T_{0,R} (2 \xi_{x}^{*}+\sinh 2 \xi_{x}^{*})}{8 \sqrt{2 a(\alpha)} 
\sigma L_{x} \cosh^{2} \xi_{x}^{*}}\, ,
\end{equation}
and the temperature $T_{0,R}$ of the gas next to the vibrating wall is
given by
\begin{equation}
\label{2.20}
T_{0,R}=\left( \frac{e}{\tanh \xi_{x}^{*}} \right)^{2}, \quad
e(\alpha)=\left( \frac{2a m}{\pi} \right)^{1/2} 
\frac{v_{b}}{\zeta^{*}}\, .
\end{equation}
The simplicity of the above results is a consequence of the limiting kind
of motion of the vibrating wall considered, whose only only effect is
to transfer energy to the grains, without inducing any periodic motion 
in the system. In some previous studies \cite{GZyB97,ByC98}, a 
``thermal'' wall, instead of a vibrating one, has been considered at $x=0$. 
By definition, particles which collide with a thermal wall leave it 
with the velocity distribution corresponding to the temperature of the wall.
Although this kind of
walls are far from reality for granular systems, their consideration might be
useful for comparison purposes. The only change to be made in the above 
discussion in order to apply it to a system with a thermal wall, 
is to replace the 
boundary condition given in Eq.\ (\ref{2.12}) by the requirement that the 
temperature $T_{0,R}$ has the value determined by the wall. As a consequence,
the expressions of the hydrodynamic profiles remain the same, while
Eq.\ (\ref{2.20}) does not apply in this case.

\section{Marginal stability analysis}
\label{s3}
Our aim now is to investigate whether the system described in the previous 
Section exhibits another steady state, in addition to the reference one. 
Then, we introduce a perturbation $\delta T(x,y)$ by
\begin{equation}
\label{3.1}
T(x,y)=T_{R}(x)+\delta T(x,y),
\end{equation}
with $\delta T(x,y) \ll T_{R}(x)$, and search for a solution of this form to 
Eqs.\ (\ref{2.9})-(\ref{2.10}) with the boundary conditions (\ref{2.11})
and (\ref{2.12}). Substitution of Eq.\ (\ref{3.1}) into Eq.\ (\ref{2.10})
and linearization in $\delta T$ yields
\begin{equation}
\label{3.2}
\left[ \frac{\partial^{2}}{\delta \xi_{x}^{2}} + 2-\frac{1}{2} \left(
\frac{\partial \ln T_{R}}{\partial \xi_{x}}  \right)^2 
+\left( \frac{\lambda_{R}}{\bar{\lambda}}\right)^2 
\frac{\partial^{2}}{\partial \xi_{y}^{2}} \right]
\delta T=\frac{4T_{R}}{p_{R}} \delta p,
\end{equation}
where $\delta p$ has been defined by $p=p_{R}+\delta p$, $\xi_{x}$ is the 
dimensionless scale introduced in Eq.\ (\ref{2.15}), and 
\begin{equation}
\label{3.3}
\xi_{y}=\sqrt{a(\alpha)}\,  \frac{y}{\bar{\lambda}}\, .
\end{equation}
Here $\bar{\lambda}$ is the average mean free path,
\begin{equation}
\label{3.4}
\bar{\lambda}=\frac{1}{2 \sqrt{2} \sigma \bar{n}}\, ,
\end{equation}
$\bar{n}=N/L_{x}L_{y}$. Next, we consider factorized solutions of Eq.\, 
(\ref{3.2}),
\begin{equation}
\label{3.5}
\delta T (\xi_{x},\xi_{y})=\phi (\xi_{x}) \varphi (\xi_{y}).
\end{equation}
The conditions at the boundaries to be verified by the functions $\phi$ and
$\varphi$ are
\begin{equation}
\label{3.6}
\left( \frac{\partial \phi}{\partial \xi_{x}} \right)_{\xi_{x}=\xi_{x}^{*}}=
\left( \frac{\partial \varphi}{\partial \xi_{y}} \right)_{\xi_{y}=0}=
\left( \frac{\partial \varphi}{\partial \xi_{y}} \right)_{\xi_{y}=\xi_{y}^{*}}
=0,
\end{equation}
\begin{equation}
\label{3.7}
\frac{1}{2} \left( \frac{\partial \ln T_{R}}{\partial \xi_{x}}
\right)_{\xi_{x}=0} \phi (0)+\left( \frac{\partial \phi}{\partial \xi_{x}}
\right)_{\xi_{x}=0}=0,
\end{equation}
where 
\begin{equation}
\label{3.7a}
\xi_{y}^{*}=2\sqrt{2a(\alpha)} \frac{\sigma N}{L_{x}}.
\end{equation}
Moreover, Eqs.\ (\ref{2.11}) and (\ref{2.12}) also imply that it must be 
$\delta p=0$, i.e. the pressure is not changed by the small perturbation.
When Eq.\ (\ref{3.5}) is substituted into
Eq.\ (\ref{3.2}), use of separation of variables leads to the equations
\begin{equation}
\label{3.8}
\frac{1}{\varphi (\xi_{y})} \frac{d^{2} \varphi (\xi_{y})}
{d \xi_{y}^{2}}=-k^{2},
\end{equation}
\begin{equation}
\label{3.9}
\left[ \frac{d^{2}}{d \xi_{x}^{2}}+2-\frac{1}{2}
\left( \frac{d \ln T_{R}}{d \xi_{x}} \right)^2  - \left(
\frac{k \bar{n}T_{R}}{p_{R}} \right)^2 \right] \phi (\xi_{x})=0,
\end{equation}
where $k$ is the constant of separation. The solution of Eq.\ (\ref{3.8}) 
verifying the corresponding boundary conditions in Eq.\ (\ref{3.6}) is
\begin{equation}
\label{3.10}
\varphi=A \cos k \xi_{y},
\end{equation}
with $A$ an arbitrary constant. Moreover, the values of $k$ are restricted to 
$k=\pi q/\xi_{y}^{*}$, $q$ being an integer. It is important to realize
that the low density limit does not imply either $\xi_{x}^{*} \ll 1$ or
$\xi_{y}^{*}\ll 1$. In fact, both quantities can be large in a very dilute
system.
An additional restriction
to be required to $\delta T(x,y)$ is that the total number of particles
in the system, $N$, is conserved. It is easily seen that this 
condition is equivalent to 
\begin{equation}
\label{3.11}
\int_{0}^{\xi_{x}^{*}} d\xi_{x}\, \int_{0}^{\xi_{y}^{*}} d\xi_{y}\, 
\frac{\delta T (\xi_{x},\xi_{y})}{T_{R}(\xi_{x})}=0\, .
\end{equation}
Equation (\ref{3.5}) with $\varphi (\xi_{y})$ given by Eq.\, (\ref{3.10})
guarantees that this equality is verified. 

An equation having a structure similar to Eq.\ (\ref{3.9}) was obtained
in ref. \cite{LMyS02} for a dense system in the limit of nearly elastic 
collisions, by employing approximate constitutive relations introduced by
Grossman {\em et al.} \cite{GZyB97}.  While the equation in \cite{LMyS02} 
was solved
using numerical techniques, here we will use a Wentzel, Kramers, Brillouin
(WKB) approximation \cite{ByO99} to investigate the possible solutions 
 \cite{note1}. First we use Eqs.\ (\ref{2.14}) and (\ref{2.19}) to
rewrite Eq.\ (\ref{3.9}) as
\begin{equation}
\label{3.12}
\frac{d^{2} \phi(\xi_{x})}{d\xi_{x}^{2}}+f(\xi_{x}) \phi (\xi_{x})=0,
\end{equation}
with
\begin{equation}
\label{3.13}
f(\xi_{x})=\frac{2}{\cosh^2 (\xi_{x}^{*}-\xi_{x})}-\frac{16 k^{2} \xi_{x}^{*2}
\cosh^4 (\xi_{x}^{*}-\xi_{x})}{(2 \xi_{x}^{*}+\sinh 2 \xi_{x}^{*})^2}\, ,
\end{equation}
which is a monotonic increasing function of $\xi_{x}$ in all the interval
$0\leq \xi_{x} \leq \xi_{x}^{*}$. 

To construct the WKB exponential approximation of Eq.\ (\ref{3.12}) it is 
necessary to consider three different ranges of parameters, that will be 
discussed separately in the following. \\*
{\em a) The function $f(\xi_{x})$ is positive everywhere in the system}. 
This is equivalent to $f(0)>0$ or 
\begin{equation}
\label{3.14}
k \leq \frac{2 \xi_{x}^{*}+ \sinh 2 \xi_{x}^{*}}{2 \sqrt{2} \xi_{x}^{*}
\cosh^3 \xi_{x}^{*}}\, .
\end{equation}
Then, the WKB solution is oscillatory,
\begin{equation}
\label{3.15}
\phi(\xi_{x})=
\frac{c_{1}}{\sqrt{g(\xi_{x})}} \exp \left[ i \int_{0}^{\xi_{x}}
d\xi_{x}^{\prime}\, g(\xi_{x}^{\prime}) \right]+
\frac{c_{2}}{\sqrt{g(\xi_{x})}} \exp \left[ -i \int_{0}^{\xi_{x}}
d\xi_{x}^{\prime}\, g(\xi_{x}^{\prime}) \right],
\end{equation}
where $c_{1}$ and $c_{2}$ are constants, and 
\begin{equation}
\label{3.16}
g(\xi_{x})=\sqrt{f(\xi_{x})}. 
\end{equation}
Imposing the boundary condition at $\xi_{x}=\xi_{x}^{*}$ leads to
\begin{equation}
\label{3.17}
\phi(\xi_{x})=\frac{c}{\sqrt{g(\xi_{x})}} \cos \int_{\xi_{x}}^{\xi_{x}^{*}}
d\xi_{x}^{\prime}\, g(\xi_{x}^{\prime}),
\end{equation}
with $c$ another constant. When the boundary condition at $\xi_{x}=0$,
Eq.\ (\ref{3.7}), is also required, the consistency condition
\begin{equation}
\label{3.18}
\tan \int_{0}^{\xi_{x}^{*}} d\xi_{x}\, g(\xi_{x})=\frac{3 \tanh 
\xi_{x}^{*}}{g^ 3(0) \cosh^{2} \xi_{x}^{*}}
\end{equation}
follows. The above equation determines the possible values of the parameters
for which a WKB solution of the differential equation (\ref{3.12}) exists
in the region under study.

If a thermal wall is considered at $x=0$, the boundary condition (\ref{3.7})
must be replaced by $\phi(0)=0$, and instead of Eq.\ (\ref{3.18}) it is
found:
\begin{equation}
\label{3.19}
\cos \int_{0}^{\xi_{x}^{*}} d\xi_{x}\, g(\xi_{x})=0,
\end{equation}
i.e.,
\begin{equation}
\label{3.20}
\int_{0}^{\xi_{x}^{*}} d \xi_{x}\, g(\xi_{x}) =\frac{(2q+1)\pi}{2},
\end{equation}
where $q$ is an arbitrary integer.  \\*
{\em b) The function $f(\xi_{x})$ is negative everywhere in the system}.
 This is  the case if $f(\xi_{x}^{*}) \leq 0$. Therefore, the region 
of parameters being considered is defined by
\begin{equation}
\label{3.21}
k \geq \frac{2\xi_{x}^{*}+\sinh 2\xi_{x}^{*}}{2 \sqrt{2} \xi_{x}^{*}}.
\end{equation}
The exponential WKB approximation in this case reads
\begin{equation}
\label{3.22}
\phi(\xi_{x})=
\frac{b_{1}}{\sqrt{h(\xi_{x})}} \exp \left[ \int_{0}^{\xi_{x}}
d\xi_{x}^{\prime}\, h(\xi_{x}^{\prime}) \right]+
\frac{b_{2}}{\sqrt{h(\xi_{x})}} \exp \left[ -\int_{0}^{\xi_{x}}
d\xi_{x}^{\prime}\, h(\xi_{x}^{\prime}) \right],
\end{equation}
with
\begin{equation}
\label{3.23}
h(\xi_{x})= \sqrt{-f(\xi_{x})},
\end{equation}
and $b_{1}$ and $b_{2}$ arbitrary constants. Imposing the boundary 
conditions (\ref{3.6}) and (\ref{3.7}) leads to the relationship
\begin{equation}
\label{3.24}
h(0)^{3} \tanh\, \int_{0}^{\xi_{x}^{*}} d\xi_{x} h(\xi_{x})
=\frac{3 \tanh \xi_{x}^{*}}{\cosh^2 \xi_{x}^{*}}\, .
\end{equation}
For a thermal wall at $x=0$, the above equation is substituted by
\begin{equation}
\label{3.25}
\cosh \int_{0}^{\xi_{x}^{*}} d\xi_{x}\, h(\xi_{x})=0,
\end{equation}
and, therefore, there is no WKB solution in this range of parameters.
\\*
{\em c) The function $f(\xi_{x})$ changes sign in the interval
$0 < \xi_{x} < \xi_{x}^{*}$}. It must be $f(0)<0$ and $f(\xi_{x}^{*})>0$. This 
corresponds to the $k$ interval
\begin{equation}
\label{3.26}
\frac{2 \xi_{x}^{*}+\sinh 2 \xi_{x}^{*}}{2 \sqrt{2} \xi_{x}^{*} 
\cosh^3 2\xi_{x}^{*}} \leq k \leq \frac{2 \xi_{x}^{*}+\sinh 2 \xi_{x}^{*}}
{ 2 \sqrt{2} \xi_{x}^{*}}.
\end{equation}
Since $f(\xi_{x})$ exhibits in this case a zero in the integration range, 
we have to consider separately the regions $\xi_{x}<a$ and $\xi_{x}>a$, 
where $a$ is the turning point, i.e. $f(a)=0$. In the former region, 
the WKB solution is given
by an expression of the form (\ref{3.22}), while in the latter it has
the form (\ref{3.15}). A global solution is constructed by matching both WKB 
approximations through the
connection formulas, expressing the connection between the oscillatory and 
exponential behaviors. Since it is easily seen that the turning point is
a simple (first-order) zero, a standard application of the theory suffices
\cite{ByO99}. Then, imposing the boundary conditions gives the following
equation to be verified by the solutions in this range:
\begin{equation}
\label{3.27}
\tan \left(\theta- \frac{\pi}{4} \right)\,  \frac{3 \tanh \xi_{x}^{*}-
h^{3}(0) \cosh^{2} \xi_{x}^{*}}{3 \tanh \xi_{x}^{*} 
+ h^{3}(0) \cosh^{2} \xi_{x}^{*}}
=\frac{1}{2} \exp \left[ -2 \int_{0}^{a} d\xi_{x} h(\xi_{x})
\right],
\end{equation}
where
\begin{equation}
\label{3.28}
\theta = \int_{a}^{\xi_{x}^{*}} d \xi_{x}\, g(\xi_{x}).
\end{equation}
The result for a thermal wall at $x=0$ is given in this case by
\begin{equation}
\label{3.29}
\tan \left( \theta -\frac{\pi}{4} \right)= \frac{1}{2} \exp \left[ -2 
\int_{0}^{a} d \xi_{x}\, h(\xi_{x}) \right].
\end{equation}

Once we have derived the equations determining the solutions to the boundary
value problem, the strategy to build up the marginal stability curve is as
follows. Given a value of $\xi_{x}^{*}$, the first question is to see whether
there is a bifurcation from the one-dimensional solution, i.e. whether
Eq.\ (\ref{3.9}) has a solution.
In case the answer to this question is affirmative, the smallest value
of $\xi_{y}^{*}$ for which there is solution gives the point of the marginal
stability curve corresponding to that value of $\xi_{x}^{*}$. For larger values
of $\xi_{y}^{*}$, the one-dimensional solution is not stable.
Suppose a solution
of Eq.\ (\ref{3.12}) exists for a given wavenumber $k$. This value is
compatible with many (infinite) values of $\xi_{y}^{*}$, the smallest
one corresponding to the choice $q=1$ in the relationship between the
possible values of $k$ and $\xi_{y}^{*}$ (see below Eq.\ (\ref{3.10})).
Moreover, the larger the value of $k$, the smaller the value of
$\xi_{y}^{*}$ associated. The conclusion is that the stability curve is
determined by the largest value of $k$ for which the problem has solution.
From a practical point of view, we have to start by
looking for solutions belonging to case {\em b} discussed above.
The values  $\xi_{y}^{*}=\pi/k$ obtained from the solutions of
Eq.\ (\ref{3.24}) are the critical values, i.e., they  define points 
on the marginal
stability curve. Nevertheless, there is no solution for every value of
$\xi_{x}^{*}$ belonging to the $k$-interval defining case {\em b}. More
precisely, there is no solution at all for a thermal wall, while it is a 
simple matter to show that the existence of solution
in this region for a vibrating wall requires that
\begin{equation}
\label{3.30}
\left[h^{(1)}(0)\right]^{3} \tanh \int_{0}^{\xi_{x}^{*}} d\xi_{x}\, 
h^{(1)}(\xi_{x})
\leq \frac{3 \tanh{\xi_{x}^{*}}}{\cosh^{2} \xi_{x}^{*}}\,,
\end{equation}
where
\begin{equation}
\label{3.31}
h^{(1)}(\xi_{x})=\lim_{k \rightarrow k_{1}} h(\xi_{x}).
\end{equation}
Equation $(\ref{3.30})$  gives $\xi_{x}^{*} \simeq 0.555$. Therefore,
solutions for larger values of $\xi_{x}^{*}$ must be found, if they exist, 
in a different region of values of $k$, namely in that considered in case 
{\em c}. With this procedure, it is an easy task to find  the largest 
value of $k$ (smallest value of $\xi_{y}^{*}$) for which the eigenvalue 
problem defined by Eq.\ (\ref{3.9}) has solution for each value of 
$\xi_{x}^{*}$, in the WKB approximation. The marginal stability curve 
for a vibrating wall obtained in this way is given in Fig.\ \ref{fig1}, 
where we have plotted
the critical values $\Delta_{c}$ of the asymmetry parameter 
$\Delta \equiv L_{y}/L_{x} =\xi_{y}^{*} / \xi_{x}^{*}$,  as a function of 
$\xi_{x}^{*}$. The solid line is the WKB solution corresponding to the
so-called case {\em b}, while the dashed line is from case {\em c}. Note
that the solutions from both regions of parameters match rather smoothly at
$\xi_{x}^{*} \simeq 0.555$. Moreover, the critical asymmetry $\Delta_{c}$ is a 
monotonic decreasing function of $\xi_{x}^{*}$, growing very fast when
$\xi_{x}^{*}$ tends to zero, as it was expected since the transversal
inhomogeneities must disappear in the elastic limit.

For a thermal wall, we have obtained
that there is no solution in region {\em b}. Therefore, we must search for
solutions belonging to the range {\em c} of parameters. Analysis of Eq.\
(\ref{3.29}) leads to the result that the equation has no solution for
$\xi_{x}^{*} \lesssim  1.181$. Moreover, Eq.\ (\ref{3.20}) has no
solution in that region either, implying that, in the WKB approximation,
the transition to the transversally inhomogeneous state requires, in the
case of system driven by a thermal wall, that the inelasticity and the number
of monolayers at rest be not small. The dotted line in Fig. \ref{fig1}
shows the marginal stability curve for a thermal wall. In the limit of
large $\xi_{x}^{*}$, the curve overlaps with the one for
a vibrated system.

\section{Simulation results}
\label{s4}
To test the above theoretical results and also to investigate the nature of 
the predicted symmetry breaking, two more fundamental descriptions of the 
system, via the direct simulation Monte Carlo (DSMC) method and 
molecular dynamics (MD) simulation, respectively, have been considered.
Although both are based on a dynamical simulation of the system, their nature 
is rather different. The DSMC method provides an algorithm to obtain numerical
solutions of the Boltzmann equation for given initial and boundary conditions
\cite{Bi94}. Therefore, it relies on the validity of a kinetic theory
description, namely the one given by the Boltzmann equation. 
On the other hand, 
no hydrodynamic approximations are introduced into the description, so that
the validity of a hydrodynamic level of description, as provided for instance 
by the Navier-Stokes equations, is not taken for granted in this approach.

The MD simulations follow the motion of the particles of the system as
a sequence of free motions and binary collisions \cite{AyT}, i.e.  
by direct application
of Newton's equations of motion. Therefore, MD provides the more basic
description of the evolution of the system. In this context, it is important
to stress that although the DSMC method also uses ``particles'' at a formal
level, they are not real particles, but fictitious ones, which are
employed to mimic the ideal dynamics described by the Boltzmann equation.
It is precisely the ideal nature of the particles in the DSMC method what
allows a very high numerical accuracy, since the number of particles
used do not affect at all the physical state being simulated. In particular,
the above number is not  related with the actual density of the
system.

In the following, we will report simulation results obtained for the system
studied in the previous sections, restricting ourselves to the vibrating wall. 
We have used the two methods, DSMC 
and MD, because they complement one another. For
instance, comparison of the results obtained by both methods provides a 
test of the validity of the inelastic Boltzmann equation to describe a
system under the physical conditions used in the MD simulations.
Since the one-dimensional state has been discussed in detail elsewhere 
\cite{BRyM00,GZyB97}, attention will
be focused here on the behavior of the system in the vicinity of the
bifurcation.

In the MD simulations, the density plays of course a relevant role. As the 
interest here is in the low density limit, small values of the surface 
fraction $\nu=N\pi \sigma^{2}/4L_{x}L{y} $, typically of the order of 
$10^{-2}$, have been used.  For fixed given values of $\nu$, $\alpha$ and 
$L_{x}$,
a set of simulations have been run corresponding to different
widths $L_{y}$ of the system, starting from a small enough value. This means
that each set of simulations corresponds to the same value of $\xi_{x}^{*}$,
as defined in Eq.\ (\ref{3.7a}), while they differ in the asymmetry parameter
$\Delta$. The simulations of the
Boltzmann equation have been carried out with a similar systematic, the
main difference being that the density plays no role in them. The other
parameter needed to specify the simulations is the velocity of the vibrating
wall $v_{b}$. We have verified that, in agreement with the theory developed in
the previous sections, the formation of transversal inhomogeneities is 
not altered by modifying this velocity, as long as it is large enough to
fluidize the complete system.

All the simulations started from a spatially homogeneous configuration, with
a Gaussian velocity distribution, corresponding to an arbitrary temperature.
The simulation is then followed until the system reaches a steady state,
in which the several monitored properties of the system 
(mean kinetic energy, density
fluctuations, and hydrodynamic profiles) become time independent. Once 
the system is in the steady state, all statistical averages of interest are
accumulated. For the purpose here, the density and temperature
profiles provide the relevant information. Let us describe what is observed 
in a set of simulations, namely we are going to present 
DSMC data from systems with $\alpha=0.95$ and
$L_{x}=10 \bar{\lambda}$. This corresponds to $\xi_{x}^{*}=1.015$. In Fig.\
\ref{fig2} the steady two-dimensional density profile for 
$L_{y}=20 \bar{\lambda}$ ($\Delta=2$) is shown in a three-dimensional plot. 
It is seen that no 
gradients in the $y$-direction are present, i.e. the system is in the 
one-dimensional reference state. The transversal homogeneity appears even
clearer in Fig.\ \ref{fig3}, where the density is plotted as a function of
$y$ for several fixed values of $x$.

When the width $L_{y}$ is increased keeping all the other parameters fixed,
a critical value shows up such that  gradients in the $y$-direction  
spontaneously  develop in the system for larger widths. 
An example of this is given in
Fig.\ \ref{fig4}, where the density surface for the same parameters as in
Fig.\ \ref{fig2}, except that now $L_{y}=26.5 \bar{\lambda}$, is plotted.
A gradient in the direction parallel to the vibrating wall is clearly
identified, becoming more pronounced next to the opposite wall, as illustrated 
in Fig.\ \ref{fig5}.
The density gradients in this case are relatively small, being the maximum 
variation of the density in the $y$ direction of the order of $5\%$. In fact,
in all the simulations, both by DSMC and MD, it has been found that when
increasing $L_{y}$, a continuous transition from the one-dimensional state
to a state with weak inhomogeneities in the transversal direction occurs.
Moreover, the transversal density profile near the transition
exhibits, as in the
case of Fig.\ \ref{fig5}, a wavelength equal to twice the width of the 
system. In the language used in Sec.\ \ref{s3}, what is observed is the
perturbation with $q=1$ ($k=\pi/\xi_{y}^{*}$), consistently with our
theoretical analysis.

If the width of the system is increased further on, the transversal 
inhomogeneities grow very fast and, of course, the results from the linear
marginal stability analysis do not apply. Simulations show that the density 
becomes very large 
in one of the corners of the system, away from the vibrating wall. The rapid 
increase of the gradients and the large value of the density in a localized
region of the system, lead to conclude that in this regime a cluster of
particles is formed \cite{LMyS02}. Of course, 
for such a region of parameters the
hydrodynamic profiles obtained from DSMC and MD are quite different, as
the former considers the particles as points while the latter assigns them
a finite diameter, implying that the density is bounded by the close-packing
value.

Once the appearance of transversal inhomogeneities has been observed by visual
inspection, it is desirable to have a ``quantitative'' criterium to
establish whether the system is or is not transversally homogeneous. 
This is equivalent to identifying an order parameter to characterize 
the transition. Since there
may be gradients in both directions, the identification of such a parameter
is not at all a trivial task. The one we have chosen is defined as follows. 
First, we introduce the dimensionless quantity $\rho_{x}(y)$ by
\begin{equation}
\label{4.1}
\rho_{x}(y)=\frac{n(x,y)}{L_{y}^{-1} \int_{0}^{L_{y}} dy\, n(x,y)}-1.
\end{equation}
If the system is transversally homogeneous, $\rho_{x}(y)$ is independent of
both $x$ and $y$, and equal to zero. When transversal inhomogeneities are
present, it depends of course on $y$ but, as it can be guessed from 
Figs. \ref{fig4} and \ref{fig5}, it also depends on $x$. 
Nevertheless, the simulation data indicate that this dependence is rather 
weak, at least near the transition. As an 
example, in Fig.\ \ref{fig6} the function $\rho_{x}(y)$ has been plotted for
the same system as in Figs.\ \ref{fig4} and \ref{fig5}. The different lines
correspond to four different values of $x$, equally separated, in the interval
$\left[ 3L_{x}/4,L_{x} \right]$. This is the region where the transversal 
gradients are larger. From the figure it follows that the $x$ dependence
is practically scaled out in the definition of $\rho_{x}(y)$. 
It must be mentioned, however,
that if the whole range of variation of $x$ is considered, some
dependence on $x$ shows up. In any case, the departure from zero 
of the average value of $\rho_{x}(y)$, $\bar{\rho}(y)$, over a certain
$x$ interval, next to the vibrating wall and not too large to avoid
$x$ dependence, provides a good criterium to distinguish transversally
inhomogeneous systems from homogeneous ones. The results to be discussed in
the following have been obtained using the interval $\left[ 3L_{x}/4,L_{x} 
\right]$.

The above discussion, the 
theoretical analysis, and the numerical results, as those illustrated in 
Fig.\ \ref{fig6}, suggest that a good order parameter may be given by the 
absolute value of the first Fourier component, $|f_{1}|$, of $\bar{\rho}(y)$.
In fact, we have verified that the absolute value of the Fourier transformed 
of this quantity exhibits an abrupt maximum for the first component when
transversal gradients begin to build up in the system.
The behavior of $|f_{1}|$ as a function of the asymmetry $\Delta$ in the 
vicinity
of the transition point, is shown in Fig.\ \ref{fig7} for the same parameters
considered in Figs.\ \ref{fig2}-\ref{fig6}. First of all, it must be noted that
$|f_{1}|$ varies in a continuous way through the transition, although it
grows very fast when one goes into the inhomogeneous region. This is the
typical behavior of the order parameter of a nonequilibrium second order 
phase transition
\cite{Ha78}. The continuous character of $|f_{1}|$ introduces some 
arbitrariness in the determination of the transition point $\Delta_{c}$. We 
have made  the choice, somewhat arbitrarily but consistently, that the 
transition takes place when $|f_{1}|$ becomes an order of magnitude
larger than its typical value in the reference state, which is determined
by the noise level. For the systems used in DSMC, this latter value is
of the order of $10^{-3}$, so that a system has been considered as
transversally inhomogeneous when $|f_{1}| \sim 10^{-2}$, which implies
deviations from homogeneity of the order of $1 \%$. This leads in the case
of Fig.\ \ref{fig6} to an estimation  $\Delta_{c}=2.5 \pm 0.1$ for the 
critical asymmetry.

By changing the initial parameters of the system and repeating the above
procedure, $\Delta_{c}$ has been computed for different values of 
$\xi_{x}^{*}$.
The results from DSMC are represented by the filled symbols in 
Fig.\ \ref{fig1}. The agreement between the theoretical predictions and the
simulation data is rather good, although a systematic deviation appears,
larger the smaller $\xi_{x}^{*}$. When valuing the comparison, it must be
taken into account that the simulations only provide an upper bound for
$\Delta_{c}$. When the system is very close to the transition point, the
time required to go from the transversally homogeneous state to the
inhomogeneous one may be too large to observe the transition during the
simulation time. In any case, it is fair to say that the hydrodynamic equations
and the WKB approximation provide an accurate description of what is observed 
in the simulations.

The results discussed up to this point were obtained from DSMC. Just to
illustrate how MD simulations lead to an equivalent scenario, we present 
next some results for a set of MD simulations with $\alpha=0.925$,
$\nu=10^{-2}$, and $L_{x}=100 \sigma$. For these values, it is
$\xi_{x}^{*}=0.281$. In Fig. \ref{fig8} a three-dimensional plot of the
density  profile is shown for $\Delta=6.4$. The system
exhibits gradients in the transversal direction, increasing again as we move 
away from the vibrating wall. However, the density gradients are very small,
being the maximum deviation from homogeneity of the order of $10 \%$. That
means that the system is close to the transition, probably in the
region where a linear approximation around the steady state still provides an 
accurate description. 

When the asymmetry is increased further on, gradients
in the perpendicular direction become sharper, and a state with a
very peaked density is observed. This is illustrated in Fig.\ \ref{fig9}
for $\Delta=7$. In Fig. \ref{fig10}, the quantity $\bar{\rho}$ is shown
for different simulations belonging to the set we are considering, i.e.
they only differ in the value of $\Delta$. The continuous transition from the 
reference state to the transversally asymmetric one is clearly observed,
as well as the dramatic increase of the transversal gradients when the
system gets well inside the unstable region. In conclusion, MD results are in
full qualitative agreement with the DSMC ones. Even more, the critical
values of the asymmetry parameters obtained from MD are in very good
quantitative agreement with those following from DSMC, as seen in Fig.\
\ref{fig1},
where they have been represented by the open symbols. This provides a
proof of the validity of the kinetic theory description as given
by the Boltzmann equation for dilute inelastic gases.

It is worth to emphasize that points in Fig.\ \ref{fig1} were obtained from
DSMC and MD simulations by changing the values of $\alpha$ and $\sigma N/L_{y}$
to sample different values of $\xi_{x}^{*}$. 
The fact that $\Delta_{c}$ obtained
in this way varies smoothly with $\xi_{x}^{*}$ supports the theoretical
prediction that the dependence on the different parameters of $\Delta_{c}$
occurs through it. This has also been ratified by considering two sets
of simulations having different values of both $\alpha$ and $\sigma N /L_{y}$,
but leading to the same value of $\xi_{x}^{*}$. The same critical asymmetry
parameter was found, supporting the scaling predicted by the theory.

\section{Discussion and concluding remarks}
\label{s5}
In this paper, the spontaneous transversal symmetry breaking in a vibrated
granular fluid predicted by Livne {\em et al.} \cite{LMyS02} has been further
investigated for low density systems. It has been shown that the
transition takes places both in MD simulations and in systems described by
the Boltzmann equation. Moreover, there is a reasonable good agreement
between the theoretical predictions, following from a marginal stability
analysis of the hydrodynamic description of the system, and the results 
from MD and also from the numerical solution of the Boltzmann equation by 
the DSMC method. This refers to the values of the critical asymmetry as
a function of the control parameter, and also to the shape of the hydrodynamic
profiles in the vicinity of the symmetry breaking.

To characterize the transition, an order parameter quantifying the  initial
set up of transversal inhomogeneities has been introduced. In terms of this
parameter, the transition presents the features of a second order
nonequilibrium phase transition. In this context, it is worth mentioning that
no sub-critical bifurcations have been observed in the simulations. Moreover,
for states well inside the instability curve, the density profile exhibits
a characteristic nonlinear $\lambda/2$ shape. On the other hand, in ref.\
\cite{LMyS02}, nonlinear two-dimensional states inside the lineal stability
region were found at high densities from the numerical solutions of the
hydrodynamic equation. Of course, there is not any contradiction 
in this, since our analysis has been  restricted to low density gases.
In ref.\ \cite{LMyS02} an analytical expression for the marginal stability
curve in a certain limit is given. In our notation, the limit considered is
$\xi_{x}^{*} \ll 1$, and the expression reads $\Delta \simeq 1.6/
\xi_{x}^{*2}$, where we have neglected subleading terms in the density.
An asymptotic analysis of the WKB results in this paper, namely of
Eq.\ (\ref{3.24}) leads to $\Delta_{c} \simeq 1.63 /\xi_{x}^{*}$, 
i.e. a qualitatively different behavior.
Given the asymptotic character of the region where these
expressions are derived, it is hard to discriminate between the two
results from the simulation data.

The work reported here shows once again the generality of spontaneous symmetry 
breaking phenomena in granular fluids. They occur in isolated as well as
in driven granular systems, in dense and dilute flows, with and without
gravitational field acting on the particles, i.e., they appear as quite a
ubiquitous effect. Interestingly, all the indications point out that
they always have a collective origin, which is fully captured by a hydrodynamic
description.

Although it may be thought that the spontaneous symmetry breaking discussed
in this paper must be closely related with the clustering instability
exhibited by a freely evolving granular gas \cite{GyZ93}, we believe this
relationship deserves more work. The information we have up to now about each
of the two  phenomena is not the same. We
know the mechanism responsible for the clustering instability (the growth of
the shear mode relative to the granular temperature), but not the
final state attracting the system. On the other hand, the existence of a
transversally inhomogeneous steady state has been established for vibrated
systems, but the detailed mechanism responsible for the development of
instabilities from the reference state is not known.

\section{Acknowledgments} This research was partially supported by the
Direcci\'{o}n General de Investigaci\'{o}n Cinet\'{\i}fica y T\'{e}cnica
(Spain) through Grant No. PB98-1124.

\begin{figure}
\caption{Marginal stability curve $\Delta_{c}(\xi_{x}^{*})$. 
The solid and dashed 
lines are the WKB predictions for a system driven by a vibrating wall, while
the dotted line is for a system with an isothermal wall. Also shown are
the results from DSMC (filled symbols) and MD (open symbols) simulations. 
For a given  value of $\xi_{x}^{*}$, the system exhibits transversal 
inhomogeneities in the steady state above the marginal curve.}
\label{fig1}
\end{figure}

\begin{figure}
\caption{Three-dimensional plot of the stationary density profile obtained by 
the DSMC method for a system with $\Delta=2$ and $\xi_{x}^{*}=1.015$. The
density is normalized by its average value $\bar{n}$, and the lengths
with the average mean free path $\bar{\lambda}$. The system does not
exhibit gradients in the $y$-direction, i.e. it is in the reference
state (bellow the marginal stability curve).}
\label{fig2}
\end{figure}

\begin{figure}
\caption{Density profiles as a function of $y$ for several fixed values 
of $x$, for the same system as in Fig.\ \protect{\ref{fig1}}. The curves
correspond, from bottom to top, to $x=L_{x}/4,
L_{x}/2, 3L_{x}/4$, and $L_{x}$.}
\label{fig3}
\end{figure}

\begin{figure}
\caption{Three-dimensional plot of the density profile obtained by the DSMC
method for the same system as in Fig.\ \protect{\ref{fig2}}, with the
only difference that now it is $\L_{y}=26.5 \bar{\lambda}$. The system is
now above the marginal stability curve and transversal gradients are
clearly observed.}
\label{fig4}
\end{figure}

\begin{figure}
\caption{Density profiles as a function of $y$ for several fixed values of
$x$ for the same system as in Fig.\  \protect{\ref{fig4}}.
The curves correspond, from bottom to top, to $x=L_{x}/4 ,
L_{x}/2 ,3L_{x}/4$ , and $L_{x}$.}
\label{fig5}
\end{figure}

\begin{figure}
\caption{DSMC results for the dimensionless function $\rho_{x}(y)$, defined 
by Eq.\ (\protect{\ref{4.1}}), for the same system as in Figs. 
\protect{\ref{fig4}} and \protect{\ref{fig5}}. The different curves 
correspond to equally spaced values of $x$ in the interval $\left[
3L_{x}/4 ,L_{x} \right]$.}
\label{fig6}
\end{figure}

\begin{figure}
\caption{DSMC results for dimensionless order parameter $|f_{1}|$ as 
a function of the asymmetry $\Delta$ for a system with $\xi_{x}^{*}=1.015$
(the dashed line has been included as a guide for the eye). 
A second order nonequilibrium bifurcation is clearly identified.}
\label{fig7}
\end{figure}

\begin{figure}
\caption{Three-dimensional plot of the stationary density profile obtained 
by MD simulation. The particles are disks of diameter $\sigma$,
the area fraction is $\nu=10^{-2}$, the coefficient of restitution
$\alpha=0.925$, $L_{x}=100 \sigma$, and $\Delta=6.4$. Small transversal
gradients are present, and the system is in the vicinity of the bifurcation.}
\label{fig8}
\end{figure}

\begin{figure}
\caption{The same as in Fig.\ \protect{\ref{fig8}} but with $\Delta=7$.
Quite large transversal gradients growing in the $x$ direction are 
identified. The system is above the marginal stability curve.}
\label{fig9}
\end{figure}

\begin{figure}
\caption{MD results for the dimensionless function $\bar{\rho}(y)$, for 
several values of the asymmetry $\Delta$, as indicated in the figure. 
All the other parameters of the
system are the same as in Figs. \protect{\ref{fig8}} and \protect{\ref{fig9}}.
The transition to a state inhomogeneous in the $y$ direction is
clearly identified.}
\label{fig10}
\end{figure}


\begin{references}
\bibitem{JNyB96} H.M. Jaeger, S.R. Nagel, and R.P. Behringer, Rev. Mod. Phys.
 {\bf 68}, 1259 (1996).
\bibitem{GyZ93} I. Goldhirsch and G. Zanetti, Phys. Rev. Lett. {\bf 70},
 1619 (1993).
\bibitem{BMGyR02} J. J. Brey, F. Moreno, R. Garc\'{\i}a-Rojo, and M. J.
 Ruiz-Montero, Phys. Rev. E {\bf 65}, 011305 (2002).
\bibitem{SyN96} H. J. Schlichting and V. Nordmeier, Math. Naturwiss. Unterr.
 {\bf 49}, 323 (1996).
\bibitem{Eg99} J. Eggers, Phys. Rev. Lett. {\bf 83}, 5322 (1999).
\bibitem{SyK01} P. Sunthar and V. Kumaran, Phys. Rev. E {\bf 64}, 041303
 (2001).
\bibitem{LMyS02} E. Livne, B. Meerson, and P. V. Sasorov, Phys. Rev. E,
 {\bf 65}, 021302 (2002).
\bibitem{BDKyS98} J. J. Brey, J. W. Dufty, C. S. Kim, and A. Santos, Phys.
 Rev. E {\bf 58}, 4638 (1998).
\bibitem{ByC01} J. J. Brey and D. Cubero, in {\em Granular Gases}, edited by 
 T. P\"{o}schel and S. Luding (Springer, Berlin, 2001), p. 59.
\bibitem{BRyM00} J. J. Brey, M. J. Ruiz--Montero, and F. Moreno, Phys. Rev. E
 {\bf 62}, 5339 (2000).
\bibitem{McyB97} S. McNamara and J-L. Barrat, Phys. Rev. E {\bf 55}, 7767
 (1997).
\bibitem{McyL98} S. McNamara and S. Luding, Phys. Rev. E {\bf 58}, 813 (1998).
\bibitem{GZyB97} E. L. Grossman, T. Zhou, and E. Ben-Naim, Phys. Rev. E
 {\bf 55} 4200 (1997).
\bibitem{ByC98} J. J. Brey and D. Cubero, Phys. Rev. E 
{\bf 57}, 2019 (1998).
\bibitem{ByO99} C. M. Bender and S. A. Orszag, {\em Advanced Mathematical
 Methods for Scientists and Engineers} (Springer, NY 1999).
\bibitem{note1} In ref. \protect{\cite{LMyS02}} a limiting case implying that
the whole system is in the dilute limit was studied analytically by means 
of a Taylor expansion. Nevertheless, the results reported differ 
significatively
from those presented here. A discussion of this point is carried out in the
final Section.
\bibitem{Bi94} G. Bird, {\em Molecular Gas Dynamics and the Direct Simulation
of Gas Flows} (Clarendon, Oxford, 1994).
\bibitem{AyT} M.P. Allen and D.J. Tildesley, {\em Computer simulations
of liquids} (Clarendon Press, Oxford, 1987). 
\bibitem{Ha78} H. Haken, {\em Synergetics: An Introduction} (Springer, Berlin,
 1978).

\end{references}
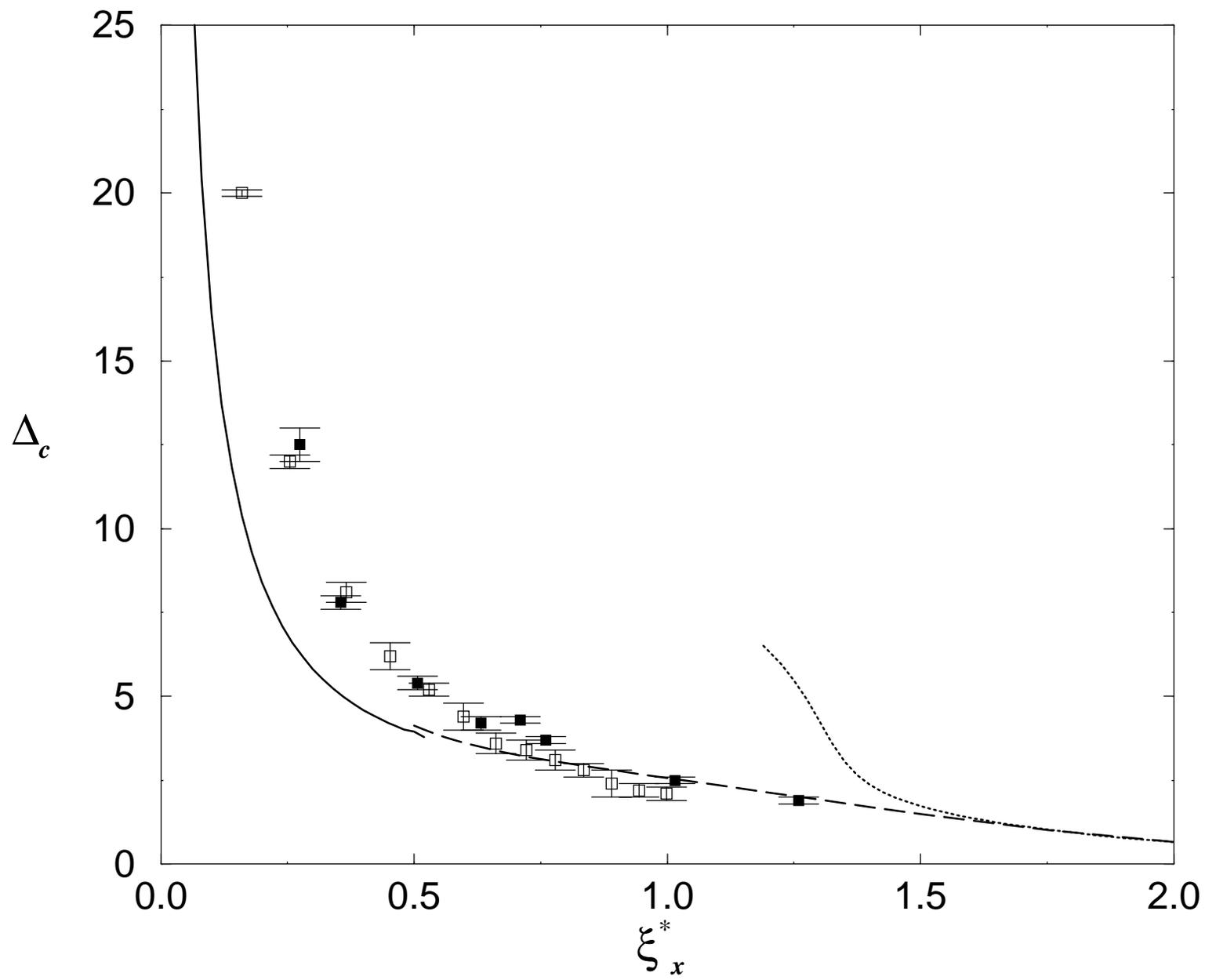
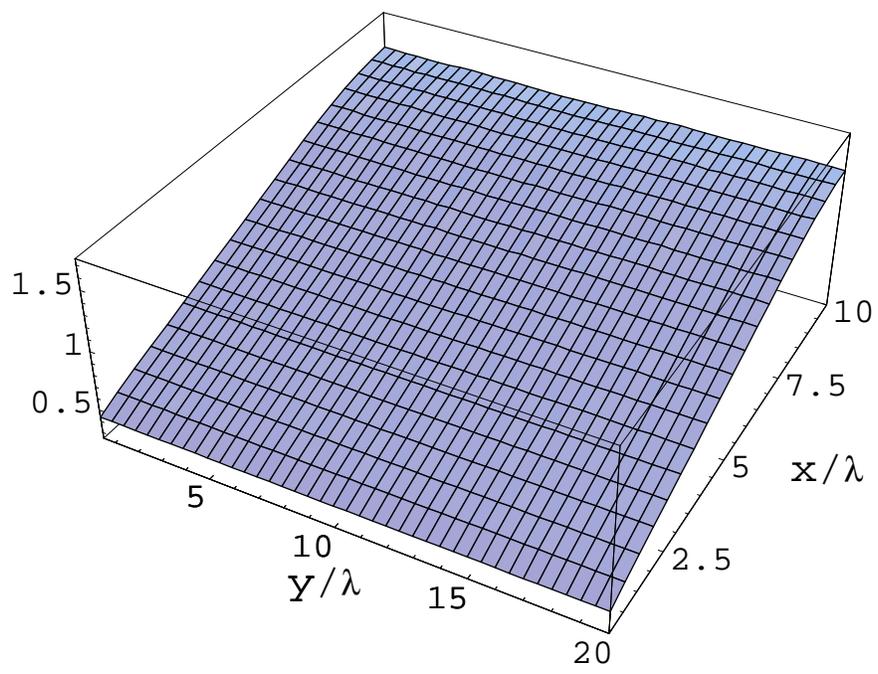
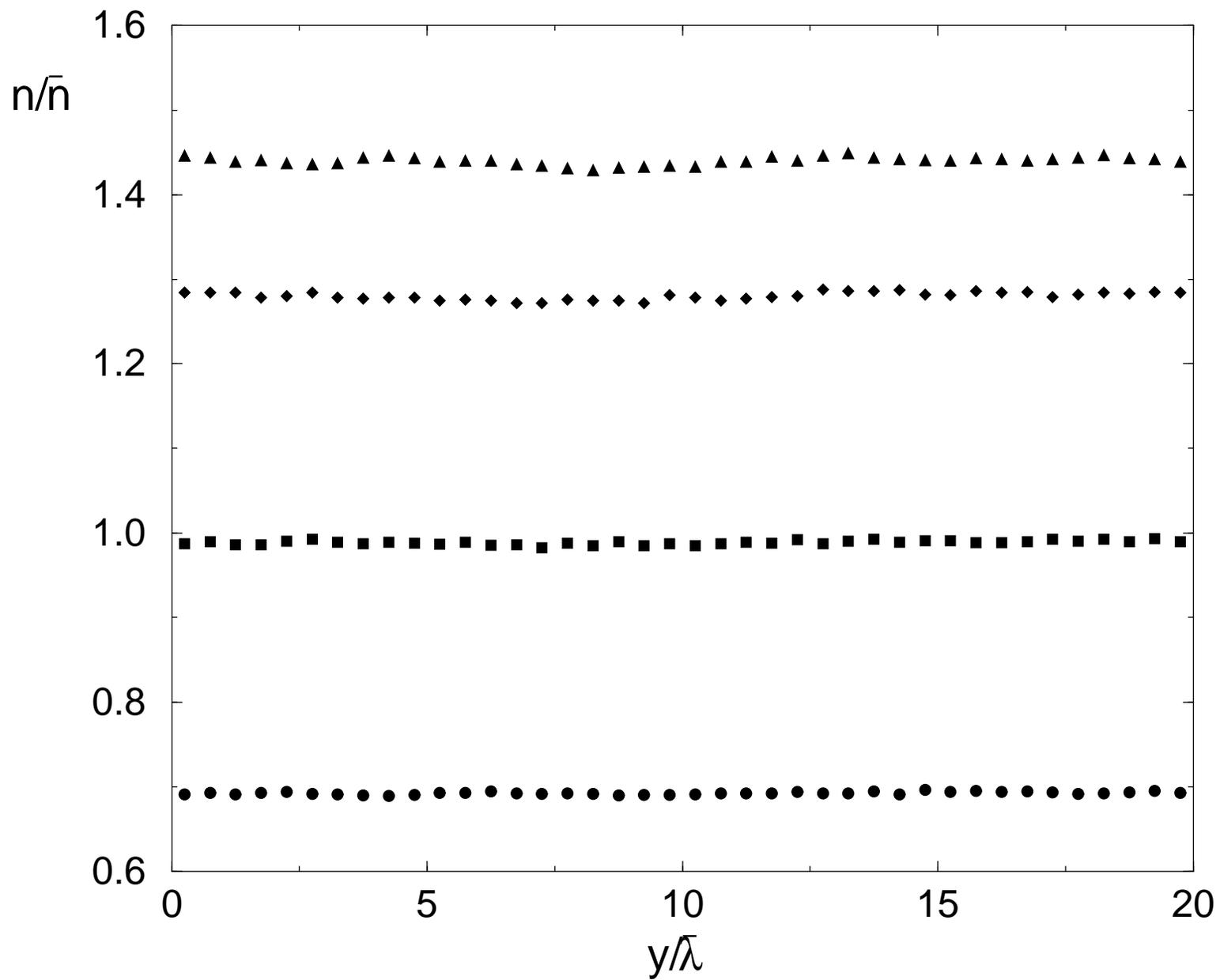
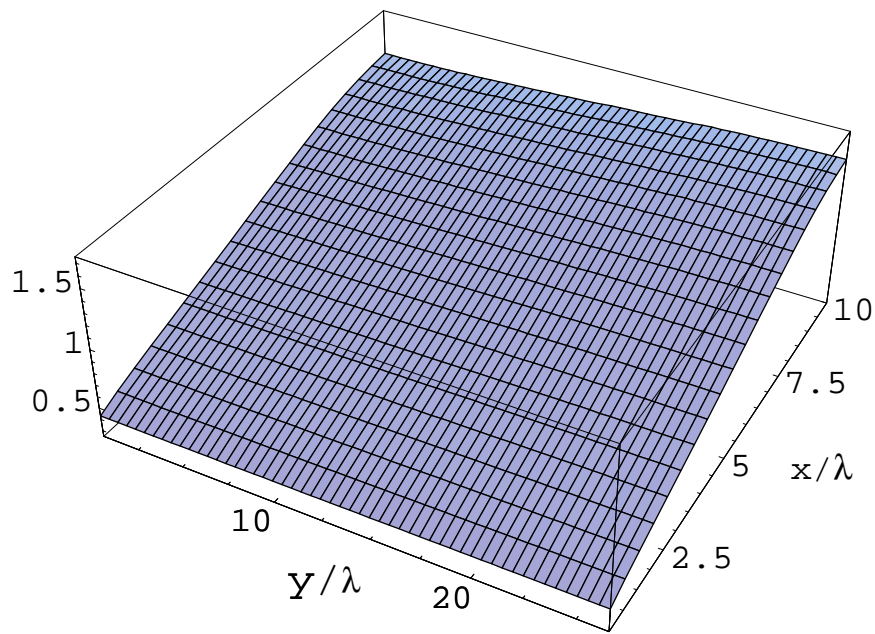
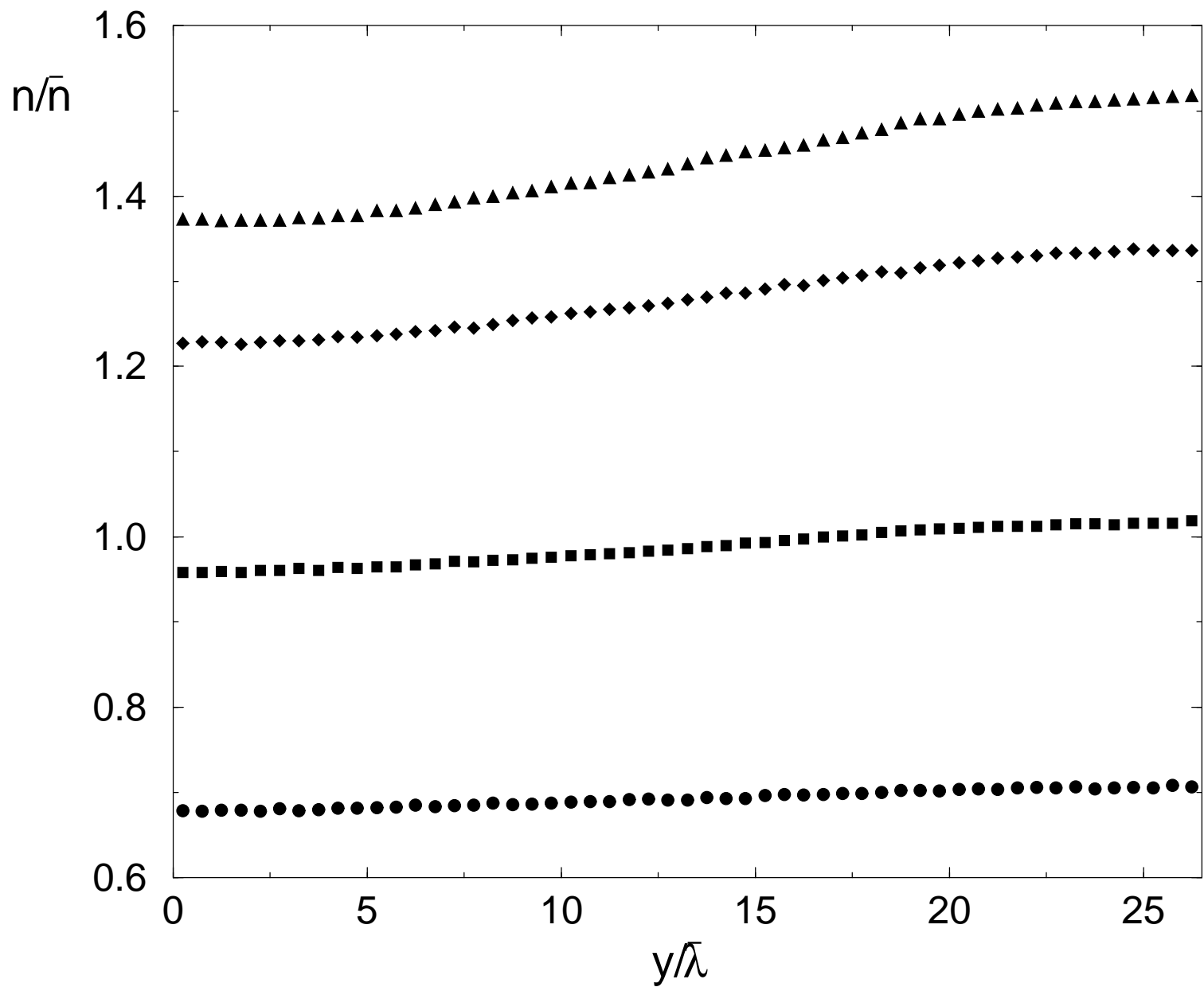
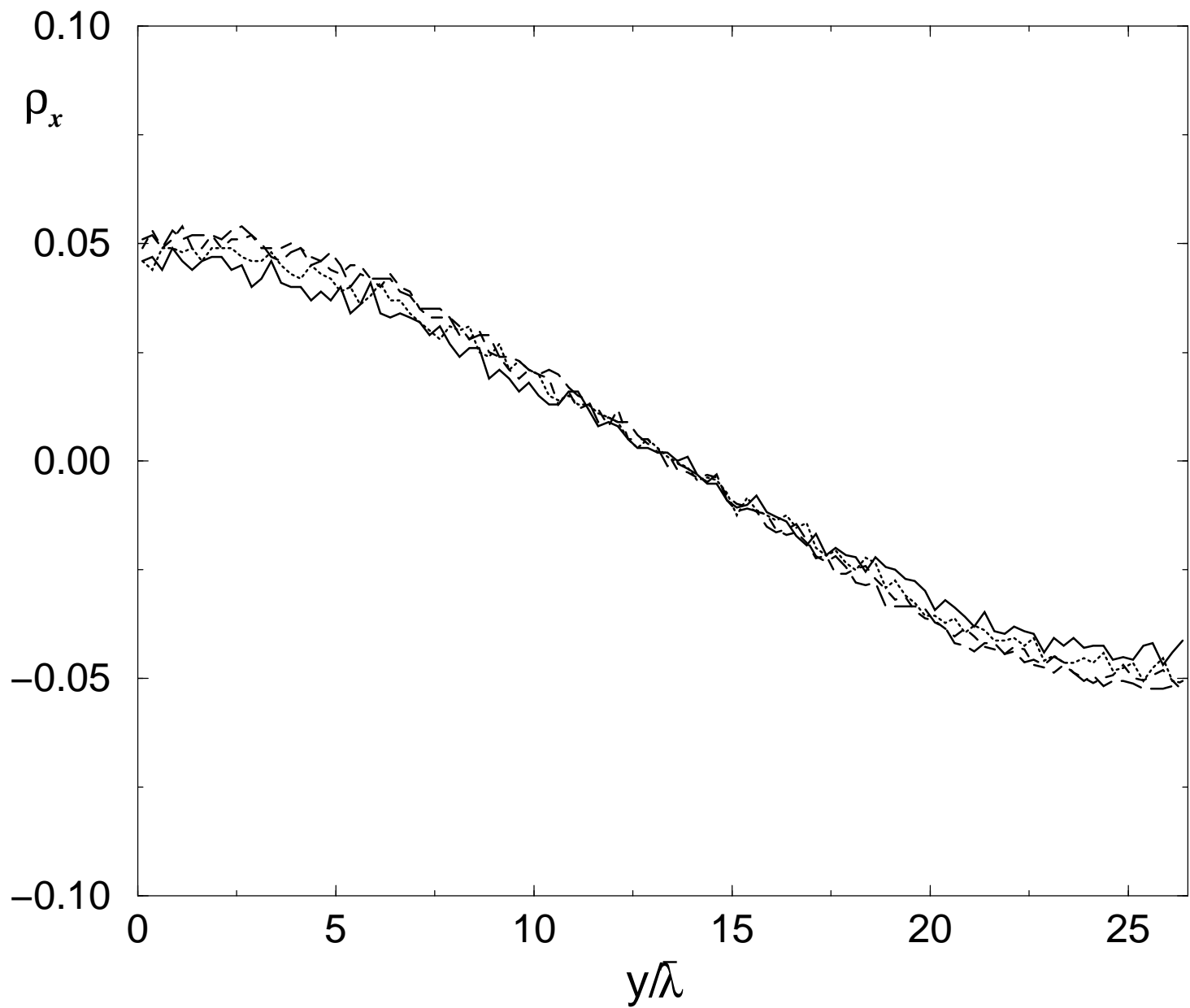
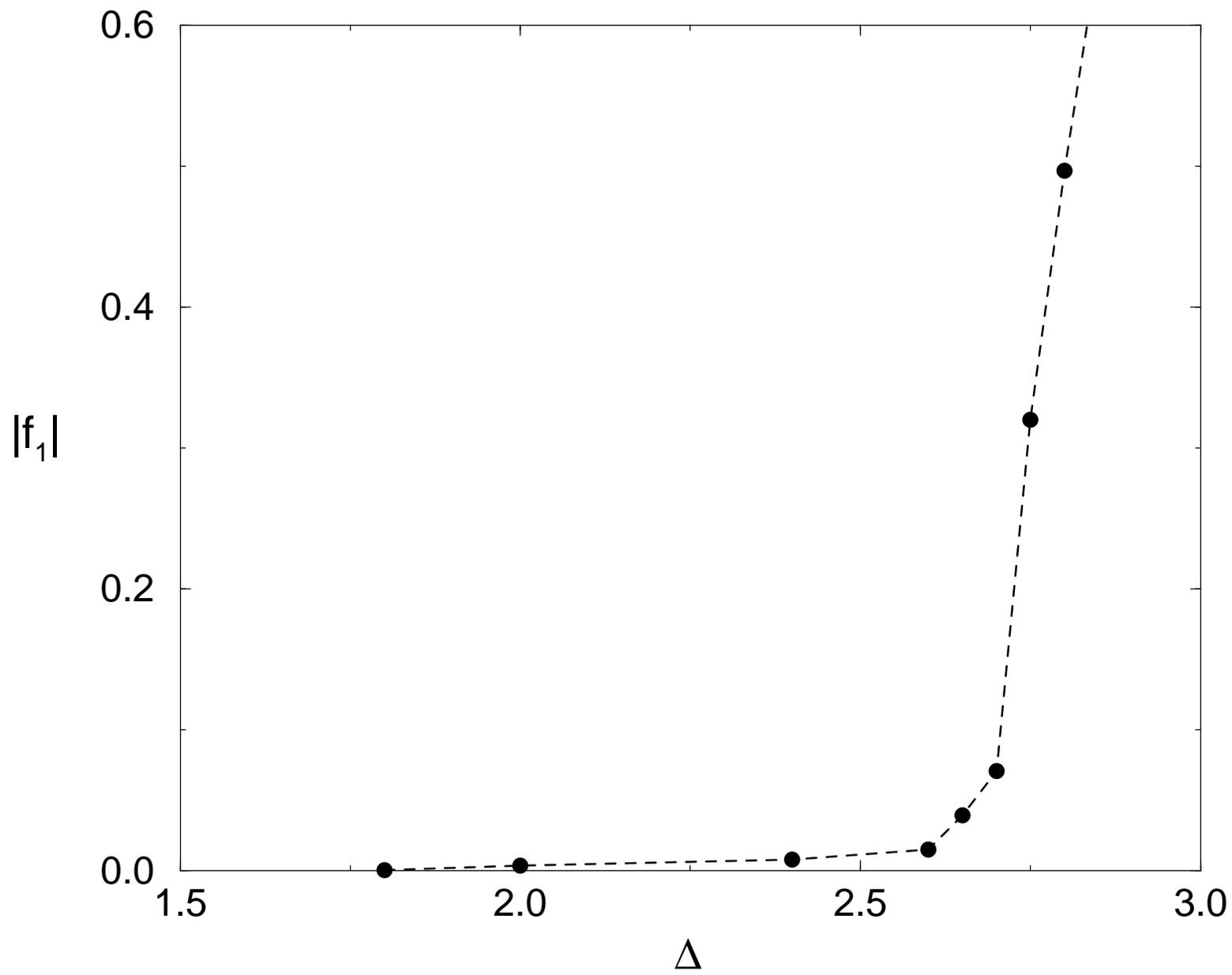
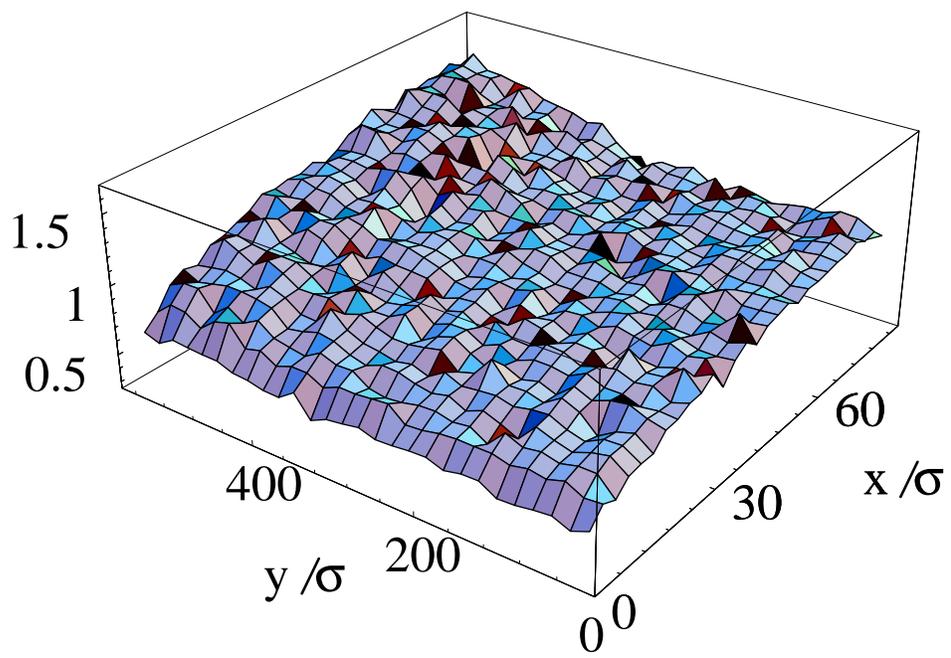
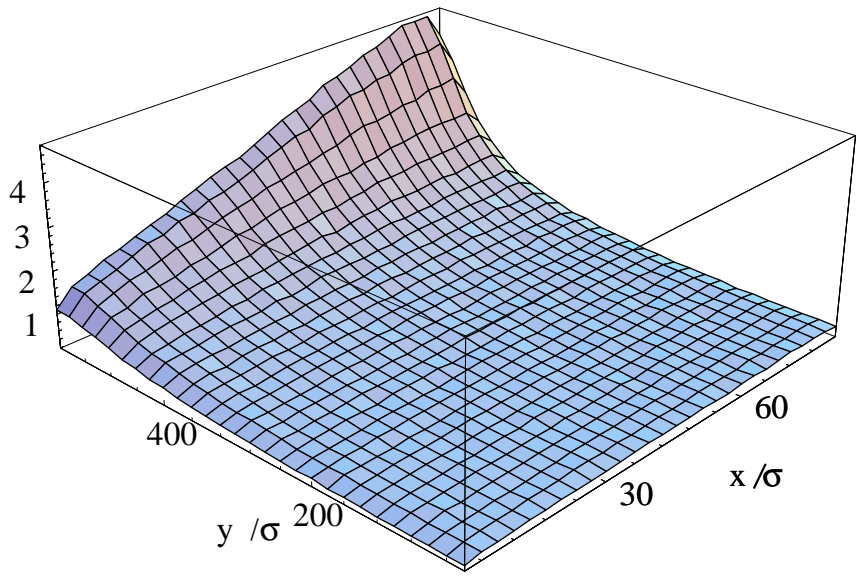
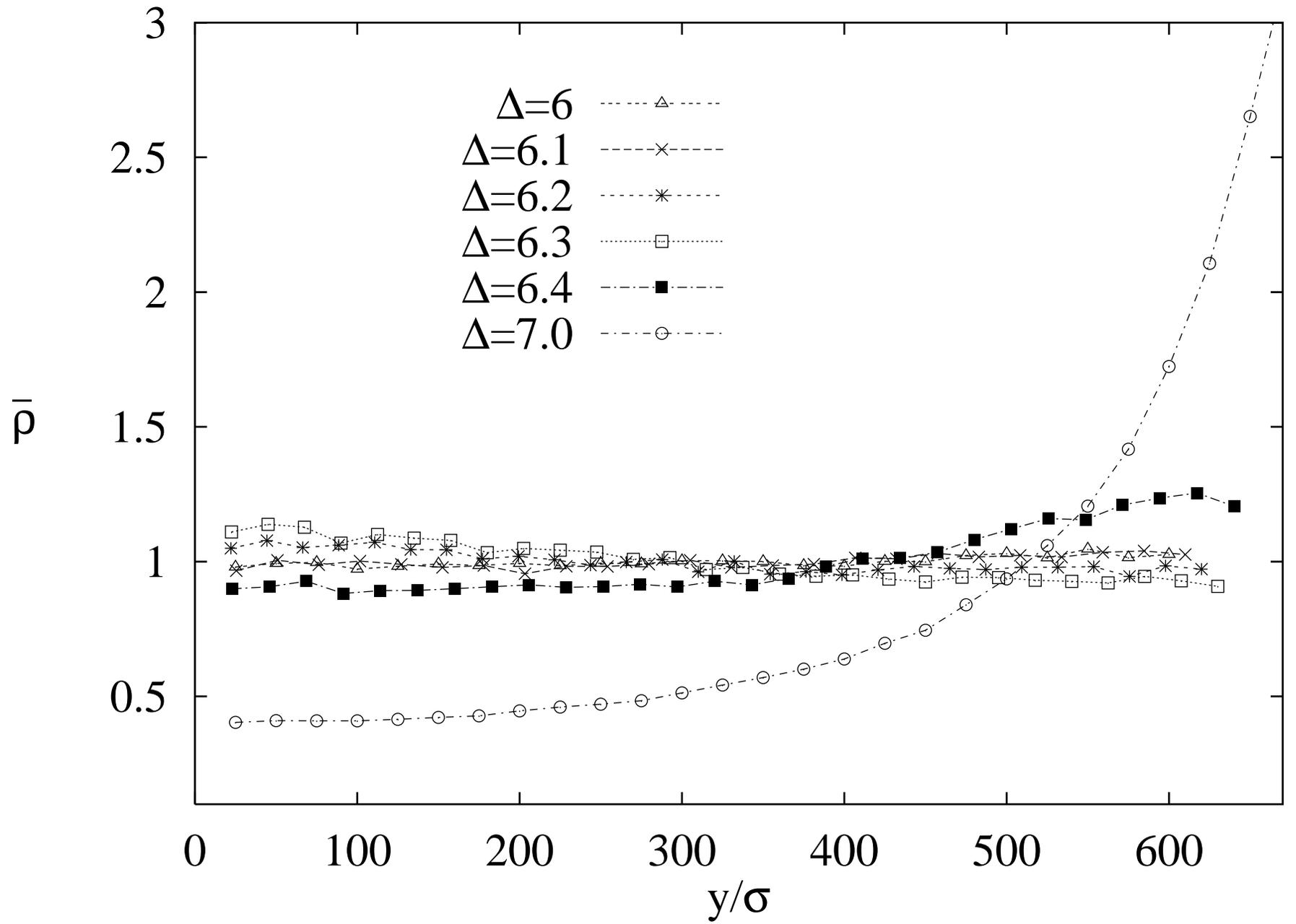
\end{document}